# The Strained State Cosmology

Angelo Tartaglia
*Politecnico di Torino*
*Italy*

## 1. Introduction

When studying cosmology one is unavoidably faced with the problem of the relevance and meaning of the terms that are in use and any purely physical and mathematical discussion borders philosophy. In this respect we must move from the remark that any description of the cosmos needs the concepts of space and time. These two entities, so fundamental in physics, are indeed neither trivial nor obvious in any respect. Going back into the past to look for the thought of the first thinkers we see for instance that Aristotle could not accept the idea of an empty space, rejecting even space as something else from the extension of existing things. "Nothing" of course does not exist, so anything in between two objects has to be something: no void, no emptiness (Aristotle, 350 b.C.).

The situation with time is even worse. The ancient Grecian thought associated time with movement and with flow, however still in the antiquity but after a few centuries we find an interesting quote from St. Augustine which gives a vivid picture of the situation: *"What is time? If nobody asks me I know, however if I wish to answer anybody asking me, I don't know"* (Augustine, 398 a.D.). I do not want to enter philosophical issues but it is wise to be aware that such fundamental questions linger in the background of any scientific discussion on cosmology.

With the birth of modern physics the question regarding the nature of space and time was posed in different terms with respect to the past, but not really solved. Newton gave definitions attributing to space and time an absolute character: an immutable stage on which physical phenomena are played within an equally immutable regular flow setting the pace for all changes and movements (Newton, 1687). This simplified and solemn view was challenged at the end of the 19th century by the failure of the Galilean transformations to guarantee the invariance of Maxwell's equations. The ether affair and the Michelson-Morley null experiment gave their contribution and finally both space and time were revisited by Einstein in his brand new Special Relativity (SR) theory. In SR length and time measurements are both observer-dependent and a new absolute entity emerges: space-time. A full description of the properties of space-time required a few years and the work of a number of scientists, not only Einstein's. At the end the relation between space-time, on one side, and matter/energy, on the other, was cast into the world famous Einstein equations:

$$G_{\mu\nu} = \kappa T_{\mu\nu} \qquad (1)$$

A problem still remained. It was and is with the nature of the left hand side of the equations. Usually space-time is thought of as a smart mathematical tool more than a physical entity, even though it interacts with matter, as the equations say. This interpretation is not explicit and some doubts remain. On the physical nature of space-time I can report a quote from a speech of Einstein's pronounced in Leiden in the 20's of the past century (Einstein, 1920):

"…. *according to the general theory of relativity space is endowed with physical qualities; in this sense, therefore, there exists an ether. … But this ether may not be thought of as endowed with the quality characteristic of ponderable media, as consisting of parts which may be tracked through time. …*"

Then space-time is real; Einstein's sentence was referred to the only space, but the implication is that the whole manifold has physical relevance even though it is not possible to treat it as matter.

That space-time is indeed something is clearly accepted by people who, since a long time and with poor results so far, are trying to quantize gravity. In these attempts space-time is often treated as a sort of field even though a subtle contradiction is implied. Fields need a background (space-time) to be described: what would the background of space-time be? Nobody has found a way out of this puzzle, at the moment.

I will not tackle directly the fundamental aspects of the problem; rather, I shall start from a simple remark. There is another branch of physics, classical physics, where a fully geometrical description is given: this is the theory of three-dimensional material continua and in particular the theory of elasticity. Even though at the beginning engineers and even physicists were not much attracted by that new mathematical language developed, at the end of the 19th century and first years of the 20th, by the Italian school (Ricci-Curbastro and Levi-Civita), after a while, thanks also to the onset of General Relativity, the whole machinery of tensor calculus was accepted. Today the elastic properties of continuous materials are currently accounted for and described in terms of tensors.

I shall elaborate on the correspondence between the general properties of space-time and the ones of ordinary material continua in order to work out a consistent description of the universe and its properties. As we shall see, the core of the theory expounded in the present chapter will be the presence in space-time of a strain energy that is the direct analogue of the elastic potential energy. The strain energy is associated with the curvature of space-time induced by the presence of matter/energy and/or by the presence of texture defects. This will be a classical approach to the other puzzling problem related with the vacuum energy. The idea of establishing a connection between a sort of rigidity of space-time and its vacuum energy is old (Sakharov, 1968), but usually implemented in terms of quantum physics and finally facing the problem of the huge mismatch between the values obtained from quantum computations and the value needed to account for the cosmological phenomena. Not all problems will be solved by this approach but many useful hints will be found.

## 2. Deformable continua

Let us start considering an N+n-dimensional space, where N and n are integers. We shall call this space *the embedding manifold* and we shall assume it is flat: the geometry in it is Euclidean. Let us cover the embedding manifold with some coordinates system that we denote with $X^a$ (*a* runs from 1 to N+n).

Next we introduce two N-dimensional embedded spaces. The first will be our *reference manifold* and is assumed to be flat; the second embedded space will be the *natural manifold* and will be intrinsically curved (Eshelby, 1956). Each embedded manifold has its own coordinates; for them I use the symbols $\xi^\mu$ (reference manifold) and $x^\mu$ (natural manifold);

the µ index runs from 1 to N. In the embedding space the reference frame is expressed by n linear constraints:

$$F_i\left(X^1,...., X^{N+n}\right) = \text{constant} \tag{2}$$

Viceversa the natural frame is fixed by n generally non-linear constraints:

$$H_i\left(X^1,...., X^{N+n}\right) = \text{constant} \tag{3}$$

The index *i* runs from 1 to n. Eq.s (2) and (3) permit to express n of the embedding coordinates in terms of the other N on the two submanifolds. In practice the N coordinates defined on each submanifold will be functions of the N+n coordinates of the embedding space: $\xi^\mu = \xi^\mu\left(X^1,...., X^{N+n}\right)$ and $x^\mu = x^\mu\left(X^1,...., X^{N+n}\right)$. For obvious convenience n will be as small as possible, i.e. in most cases it will be n = 1; however for peculiar natural frames containing singularities one more dimension can be insufficient to give a flat embedding, so more will be required.

As an additional assumption, suppose, for the moment, that the natural manifold is sufficiently regular and all functional dependences are smooth and differentiable as many times as needed. As a consequence it will be possible to directly express the coordinates on the reference manifold as functions of those on the natural manifold and viceversa.

Once the above definitions and conditions have been declared we may establish a one to one correspondence between points located on the two embedded manifolds. This correspondence is embodied in an **u** vector field: each **u** vector goes from a point in the reference to a point in the natural manifold. The flatness of the embedding space permits a global definition of the vector field. The situation described so far is summarized in fig. 1. The vector **u** field is called the *displacement vector field*; whenever it is non-uniform we say that the natural manifold is distorted with respect to the reference one.

Considering pairs of arbitrarily near positions on both manifolds we may compare the corresponding line elements. Let us write

$$d\sigma^2 = \eta_{\mu\nu} d\xi^\mu d\xi^\nu \tag{4}$$

for the reference manifold. Due to the flatness condition it must also be

$$\eta_{\mu\nu} = \delta_{\alpha\beta} \frac{\partial y^\alpha}{\partial \xi^\mu} \frac{\partial y^\beta}{\partial \xi^\nu} \tag{5}$$

The y's are Cartesian coordinates and the metric tensor $\eta_{\mu\nu}$ corresponds to an Euclidean geometry in N dimensions.

For the natural manifold it will be

$$ds^2 = g_{\mu\nu} dx^\mu dx^\nu \tag{6}$$

Both line elements (5) and (6) can of course be expressed in the embedding space as

$$ds^2 = \delta_{ab} dX^a dX^b \tag{7}$$

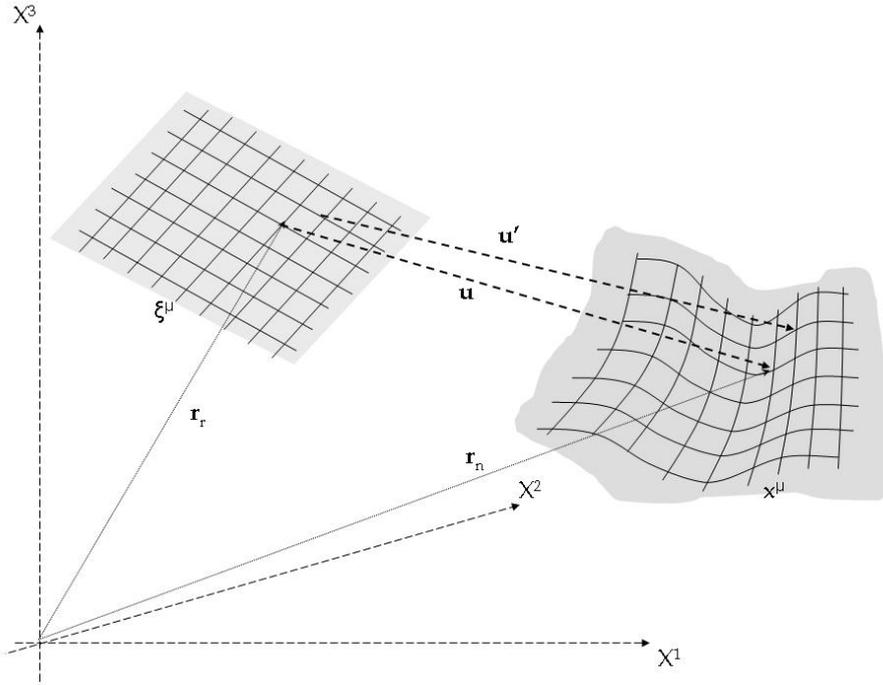

Fig. 1. The embedding space with the two embedded manifolds. The figure represents a three-dimensional embedding of two bidimensional manifolds, but the scheme can be applied to any number of dimensions.

where Cartesian coordinates are assumed, for simplicity; Latin indices from the first part of the alphabet (as a, b, c…) run from 1 to N+n. One goes from (7) to (5) or (6) applying respectively the constraints (2) and (3) and remarking that (see fig. 1) it is:

$$\mathbf{r}_n = \mathbf{r}_r + \mathbf{u} \tag{8}$$

Summing up and using (8) we see that the difference between (6) and (4) is:

$$ds^2 - d\sigma^2 = \delta_{\mu a} \frac{\partial u^a}{\partial x^\nu} + \delta_{\nu a} \frac{\partial u^a}{\partial x^\mu} + \delta_{ab} \frac{\partial u^a}{\partial x^\mu} \frac{\partial u^b}{\partial x^\nu} \tag{9}$$

The difference (9) has been written in terms of the coordinates on the natural manifold. Using on both sides the same coordinates, eq. (9), together with (4) and (6), leads to:

$$g_{\mu\nu} - \eta_{\mu\nu} = 2\varepsilon_{\mu\nu} \tag{10}$$

The elements $\varepsilon_{\mu\nu}$ belong to a rank 2 symmetric tensor in N dimensions: it is called the *strain tensor*.

So far the correspondences we have established may be though of as being purely formal, however if we consider a physical situation we may think of obtaining the natural manifold from the reference one by continuous deformation. In this case the displacement vector tells us from where to where a given point has been moved during the process and the differential part of the displacement does indeed represent the strain induced in the manifold.

## 2.1 Defects

The conceptual framework outlined in the previous section permits to introduce another important notion: the one of *defect* or *texture defect*.

Defects play an important role in the analysis of the properties of crystals or, in general, of material continua. A consistent description for them was worked out between the end of the 19th and the beginning of the 20th century (Volterra, 1904) and that is the picture I shall use in the following.

Consider the situation represented in fig. 2, whose general structure is the same as that of fig.1. We say we have a defect whenever a whole region $\mathcal{C}$ of the reference manifold corresponds to a point O (or a line or any other lower dimensional subset) of the natural manifold, while, for the rest, the correspondence remains one to one.

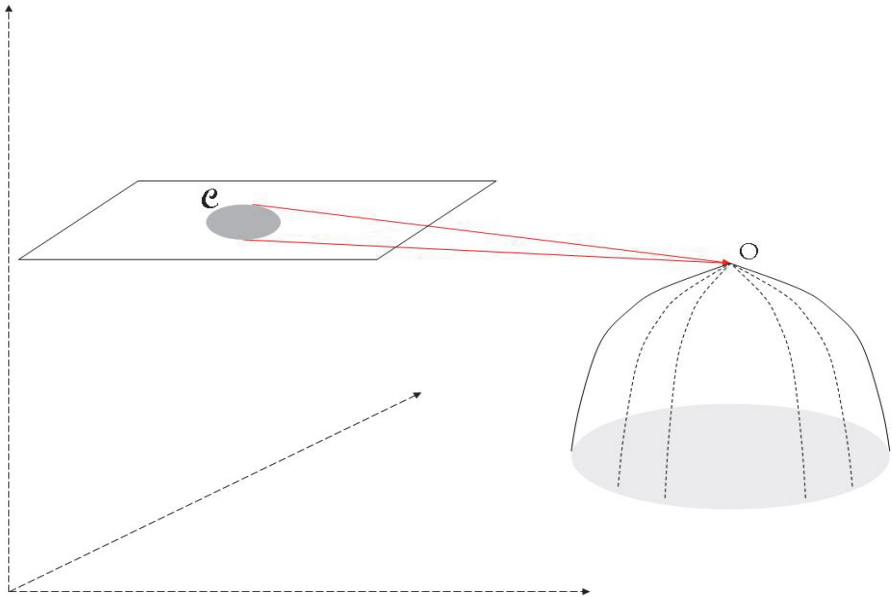

Fig. 2. Defects in continuous manifolds. Point O corresponds to a whole region $\mathcal{C}$ of the reference manifold. The natural manifold has non-zero strain.

The presence of a defect implies a non-zero strain tensor in the natural manifold and the strain is singular in correspondence of the defect. Defects also induce peculiar symmetries in the natural manifold: a pointlike defect induces a central (spherical) symmetry; a straight linear defect implies a cylindrical symmetry, etc. A whole classification of defects, on the basis of the corresponding symmetries, exists in terms of dislocations and disclinations. Volterra's classification has been extended to space-time by Puntigam and Soleng (Puntigam, 1997) who identified the 10 possible types of distortions existing in four dimensions; they wanted to apply the idea of topological defects to the study of cosmic strings. I will not enter into further details, since the general concepts are enough for the purpose of this chapter.

## 2.2 Elasticity

In physical terms, strain is not enough to account for what happens. We must say something about the causes of the distortion of the manifold and their interrelation with the effects. In other words, when we try to deform a material system (the reference manifold of our abstract representation) we expect it to react back to our action. In three dimensions the reaction is in term of stresses in the bulk of the material: strains are relative changes in the linear sizes; stresses are forces per unit surface and altogether they form the rank 2 symmetric stress tensor, $\sigma_{\mu\nu}$. Stresses and strains are mutually and causally connected to each other; in this connection consists the *elasticity* of the material. The simplest assumption we can make is that the relation between strain and stress is linear. Indeed if we exclude discontinuities in the behaviour of the continuum we are analyzing, linearity is in any case the lowest order approximation for the strain/stress functional dependence. Let us then limit our study to the linear elasticity case; its basic equation is Hooke's law, which, in tensor notation, is written:

$$\sigma_{\mu\nu} = C_{\mu\nu}{}^{\alpha\beta} \varepsilon_{\alpha\beta} \tag{11}$$

The $C_{\mu\nu}{}^{\alpha\beta}$'s are the elements of a rank 4 completely symmetric tensor, which we can call the elastic modulus tensor; it contains the properties of the material at the linear approximation level. Eq. (11) is a tensor equation so it is covariant and locally coinciding with its expression on the tangent space; this means that the upper or lower position of the indices is simply a matter of convenience in order to exploit Einstein's summation convention[1].

If we assume that our material continuum is locally isotropic, simple symmetry arguments tell us that the elastic modulus tensor only depends on two parameters, known as the Lamé coefficients, $\lambda$ and $\mu$, of the material. Explicitly one has:

$$C_{\alpha\beta\mu\nu} = \lambda \eta_{\alpha\beta} \eta_{\mu\nu} + \mu \left( \eta_{\alpha\mu} \eta_{\beta\nu} + \eta_{\alpha\nu} \eta_{\beta\mu} \right) \tag{12}$$

Eq. (12) is written for an arbitrary choice of the coordinates; using Cartesian coordinates the $\eta$'s would be replaced by Kronecker $\delta$'s. Using (12) Hooke's law becomes:

$$\sigma_{\mu\nu} = \lambda \varepsilon^{\alpha}{}_{\alpha} \eta_{\mu\nu} + 2\mu \varepsilon_{\mu\nu} \tag{13}$$

### 2.2.1 Deformation energy

It is convenient to write down the elastic potential energy of the strained state, which is $W = \frac{1}{2} \sigma_{\mu\nu} \varepsilon^{\mu\nu}$. Using eq. (13) we obtain:

$$W = \frac{1}{2} \lambda \varepsilon^2 + \mu \varepsilon_{\mu\nu} \varepsilon^{\mu\nu} \tag{14}$$

---

[1] Some care will be required when treating a manifold with Lorentzian signature.

Now I have posed the trace of the strain tensor $\varepsilon^\alpha_{\ \alpha} = \varepsilon$ for short.
Eq. (14) could have been written also considering the lowest significant terms of the Helmholtz free energy $F_H$ of the material, written in terms of strain. In fact $F_H$ must contain only scalar quantities and, besides a constant, its lowest order is the second, because the thermodynamical equilibrium must correspond to a minimum (Landau, 1986). Eq. (14) contains the only two second order scalars that can be built from the strain tensor.

## 3. Space-time and the universe

The whole description of strained continua is molded on three-dimensional examples, but the treatment holds for any number of dimensions. Of course one needs to generalize the interpretation of such things as the stresses and the energy, but formulae and criteria remain valid. So let us apply the theory to four dimensions and the Lorentzian signature, i.e. to space-time, treated as a physical continuum endowed with properties analogous to the ones of ordinary elastic materials.

As a first step I will generalize the action integral of space time plus matter/energy. The generalization consists in that a strained state is associated with a potential like the one expressed in eq. (14). The additional term will appear in the Einstein-Hilbert action that becomes:

$$S = \int \left( R + \frac{1}{2}\lambda\varepsilon^2 + \mu\varepsilon_{\mu\nu}\varepsilon^{\mu\nu} + \mathcal{L}_{mat} \right)\sqrt{-g}\,d^4x \tag{15}$$

Now the scalar curvature $R$ plays the role of dynamical term, since it contains the derivatives of the Lagrangian coordinates, i.e. the elements of the metric tensor; $\mathcal{L}_{mat}$ is the Lagrangian density of matter/energy. Eq. (15) is the starting point for what I shall call the Strained State Theory (SST), which in the following will be applied to the Strained State Cosmology (SSC).

From (15) we can also derive generalized Einstein equations. The new elastic potential terms contribute an additional stress/energy tensor in the final equations. We may treat the strain tensor in the same way as we do with matter fields, only remembering that it must satisfy the constraint represented by eq. (10). In particular the indices of the strain tensor are raised and lowered using the full metric tensor. On this footing we obtain the new generalized version of eq. (1) in the form:

$$G_{\mu\nu} = T_{(e)\mu\nu} + \kappa T_{\mu\nu} \tag{16}$$

In explicit form it is:

$$T_{(e)\mu\nu} = \lambda\varepsilon\varepsilon_{\mu\nu} + 2\mu\varepsilon_{\mu\nu} \tag{17}$$

The tensor $T_{(e)\mu\nu}$ actually belongs to space-time (it is in a sense a self-interaction energy) but works as an effective additional term on the side of the sources.

### 3.1 A Robertson-Walker universe

It is commonly assumed that the universe has a Robertson-Walker (RW) symmetry, i.e. it is homogeneous and isotropic in space (cosmological principle). This conviction is based both on a priori arguments and on the observation. On the theoretical side: why should a given position or direction in space be more important than another? So let us assume all positions and directions are equivalent. In the 19th and at the beginning of the 20th century, as well as later on, at the time of the Hoyle-Gold-Bondi steady state cosmology, this argument was assumed to hold also for time: why should any given moment be "special"? The homogeneity of time together with the homogeneity and isotropy of space forms the so called "perfect cosmological principle".

The four-dimensional homogeneity has however almost completely been abandoned on the basis of observation. Strictly speaking a stationary universe had already been challenged by the Olbers' paradox (1826): why is the sky dark at night? However the crucial data came from Hubble's work at the end of the 20's of the last century: the redshift of the light coming from other galaxies tells us that the universe is expanding. Today, after the publication of the observations by the groups led by Adam Riess (Riess, 1998) and Saul Perlmutter (Perlmutter, 1999), we even think that the expansion of the universe is accelerating.

As for the homogeneity and isotropy of space the observational evidence is not so stringent. It is evident that locally the universe is neither homogeneous nor isotropic. One has to go to a large enough scale to override local inhomogeneities and anisotropies; how large? Actually we see large voids in the universe, then huge filaments made of galaxies, so that the cosmological principle is assumed to hold at a scale of at least hundreds of megaparsecs (Mpc). However it is also true that we have knowledge only of the visible part of the universe; of the rest we cannot say almost anything or even nothing at all. In fact various anisotropic solutions for the Einstein equations applied to cosmology have been studied and the possibility that some "local" inhomogeneity is responsible for what has been interpreted as an accelerated expansion has also been considered (Biswas, 2010).

I will not discuss further these issues, but will stay with the standard cosmology and accept that the cosmological principle holds on the average. This assumption greatly simplifies the discussion of the global behaviour of the universe and is synthetically expressed by the Robertson-Walker symmetry.

A question is however legitimate now: why is the RW symmetry there? If you just add a uniformly distributed dust to an empty Minkowskian space-time you do not obtain, as an unique outcome, a RW universe. A homogeneous distribution of matter is gravitationally unstable; does this preserve isotropy and lead to a singularity in the past? Not necessarily.

If I adopt the viewpoint of the SSC, I may think that space-time *per se* (the natural manifold) has a built-in RW symmetry independently from the presence of matter; the latter simply responds to the symmetry, reinforcing it. The primordial symmetry is in turn explained assuming the presence of a spacelike defect (a Cosmic Defect) within the manifold. Of course you might ask why the defect should be there, however we know that going back along the chain of "why?"'s sooner or later we exit the domain of physics. We can only try and minimizing the number of independent assumptions and if possible look for physically consistent interpretations of their meaning.

The approach of the Strained State Cosmology is best visualized in fig. 3, where the embedding of a Robertson-Walker space-time in a three-dimensional flat manifold is shown. $O$ is the defect responsible for the RW symmetry. For convenience in making the drawing, the example of a closed space has been represented. For an open space the original defect would be linear (a ridge) and space-like. All geodetic lines starting from the defect are time-like; $\tau$ is the cosmic time; space is any space-like intersection between the natural manifold and an open surface (for instance a hyperplane) in the embedding space. Successive intersections of the natural manifold, in correspondence of increasing values of the cosmic time, evidence what the typical 3+1 human view reads as an expanding universe.

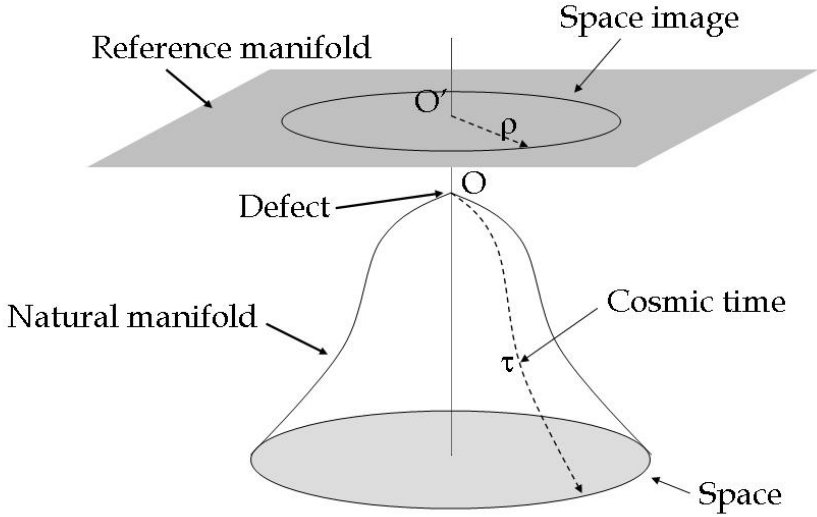

Fig. 3. Pictorial view of a Robertson-Walker universe embedded in a three-dimensional flat space. The picture corresponds to a closed universe.

The correspondence we establish between the reference and the natural manifold identifies an "image" of any given natural space in the reference. We must now write down and compare the corresponding line elements on the two manifolds. Due to the simple symmetry, the line element on the natural manifold is of course[2]:

$$ds^2 = d\tau^2 - a^2(\tau)\, dl^2 \tag{18}$$

The $a$ function of the cosmic time is the *scale factor* of the universe; dl is the space length element.

As for the reference manifold you can in principle define the correspondence with the actual RW space-time in infinite different ways. Using the coordinates chosen for the natural manifold, you are left with four free functions for the choice of the coordinates on the reference, with the constraint that the reference has to be flat. In the specific case under consideration, however, the final symmetry reduces the free functions to only one and the reference line element is written:

$$d\sigma^2 = b^2(\tau)\, d\tau^2 + dl^2 \tag{19}$$

The function $b$ of the cosmic time has been called gauge function in (Radicella, 2011) but this denomination is not entirely correct, since $b$ does not correspond to a real freedom: since we assume that the deformation process is a real one, the way the correspondence between the unstrained and the strained manifold is established depends on the two Lamé coefficients of space-time, under the assumption of local isotropy.

---

[2] Times are expressed as lengths.

From eq.s (18) and (19), using the definition (10), we easily obtain the non-zero elements of the strain tensor for a RW space-time:

$$\begin{cases} \varepsilon_{oo} = \dfrac{1-b^2}{2} \\ \varepsilon_{ii} = -\dfrac{1+a^2}{2} \end{cases} \tag{20}$$

Once we have the strain tensor, it is possible to deduce the potential term (14) in the action integral; indices are raised and lowered by means of the full RW metric tensor. It is:

$$W = \frac{\lambda}{8}\left(1 - b^2 + 3\frac{1+a^2}{a^2}\right)^2 + \frac{\mu}{4}\left[\left(1-b^2\right)^2 + 3\frac{\left(1+a^2\right)^2}{a^4}\right] \tag{21}$$

The other ingredients of the action integral, besides the matter/energy Lagrangian density, are:

$$R = -6\left(\frac{\ddot{a}}{a} + \frac{\dot{a}^2}{a^2}\right); \qquad \sqrt{-g} = a^3 \tag{22}$$

Dots denote derivatives with respect to time.

An expression for $b^2$ is immediately found imposing $dW/db = 0$ (i.e. extremizing the Lagrangian density with respect to the gauge function). Rejecting the inadmissible $b = 0$, the solution is:

$$b^2 = 2\frac{2\lambda + \mu}{\lambda + 2\mu} + \frac{3}{a^2}\frac{\lambda}{\lambda + 2\mu} \tag{23}$$

Given the solution (23) the only residual unknown is the scale factor $a$. Of course we should also specify the type of matter we consider. The simplest is to assume that matter/energy is made of dust plus radiation. Under these conditions, applying Hamilton's principle to the action integral (15) leads to:

$$H = \frac{\dot{a}}{a} = c\sqrt{\frac{3}{16}B\left(1 - \frac{(1+z)^2}{a_0^2}\right)^2 + \frac{\kappa}{6}(1+z)^3\left[\rho_{m0} + \rho_{r0}(1+z)\right]} \tag{24}$$

$H$ is the Hubble parameter. The variable $z$ is the redshift factor and use has been made of the relation $a(1+z)$ = constant = $a_0$; $a_0$ is the present value of the scale factor. $\rho_{m0}$ and $\rho_{r0}$ are the present values of the average matter and radiation densities in the universe; $\kappa = 16\pi G/c^2$ is the coupling constant between geometry and matter/energy. $B$ combines the Lamé coefficients of space-time according to:

$$B = \frac{3}{2}\mu\frac{2\lambda + \mu}{\lambda + 2\mu} \tag{25}$$

The term proportional to *B* in the square root of eq. (24) is the contribution coming from the strain of the space-time; the rest is the standard cosmology of a RW universe filled up with dust and radiation.

The choice of the sign for the square root in (24) tells us whether the universe is expanding or contracting; the given behaviour is for ever. In the same time we see that the contribution from strain implies the onset of acceleration after an initial phase of deceleration. The dependence of the expansion rate on the scale factor is shown in fig. 4 in arbitrary units. At very early times ($z \gg 1$) the strain contributes a radiation-like term boosting the expansion:

$$H_{z \gg 1} \cong cz^2 \sqrt{\frac{3}{16}\frac{B}{a_0^4} + \frac{\kappa}{6}\rho_{r0}} \qquad (26)$$

In late times ($z \to -1$) the Hubble parameter becomes constant: the expansion assumes an exponential trend at a rate depending only on *B*:

$$H_{z \to -1} \cong c\sqrt{\frac{3}{16}B}; \qquad a_\infty \approx e^{c\sqrt{\frac{3}{16}B}\,t} \qquad (27)$$

We have so seen that the SSC is able to account for the accelerated expansion as being a consequence both of the presence of a cosmic defect (the Big Bang) and of the elastic properties of space-time.

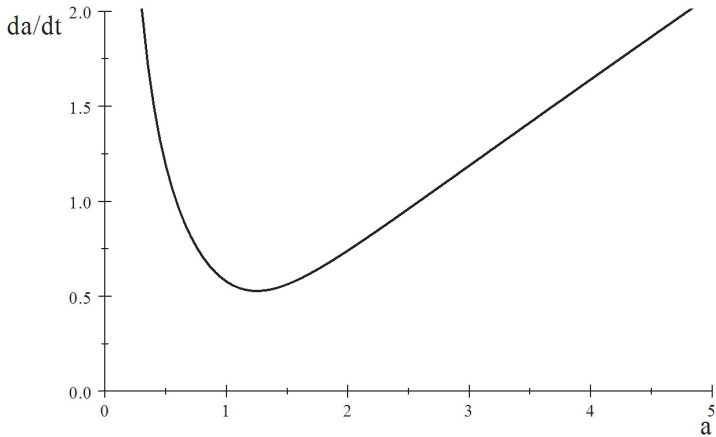

Fig. 4. Expansion rate of a RW universe according to the Strained State Theory. The graph is drawn giving arbitrary values to the parameters. The universe always expands; at the beginning the expansion decelerates, afterwards it accelerates.

What remains to be done is to find appropriate values for the parameters of the theory, which, at this stage, are *B* and $a_0$ besides $\rho_{m0}$ and $\rho_{r0}$. This will be the subject of the next section.

## 4. Cosmological tests

In order to determine the optimal values for the parameters of the theory and to check its credibility we have considered four typical tests: the dependence of the luminosity of type Ia supernovae (SnIa) on the redshift; the Big Bang Nucleosynthesis (BBN); the acoustic horizon scale in the Cosmic Microwave Background; the Large Scale Structure (LSS) formation after the recombination era. The first test I have quoted is not in decreasing redshift order as the others are; the reason for privileging it is in that SnIa's have been the first evidence in favor of an accelerated expansion (Riess, 1998) (Perlmutter, 1999).

### 4.1 The luminosity curve of type Ia supernovae

Type Ia supernovae are thought to be the product of the implosion of a slowly rotating white dwarf star that accretes matter from a companion in a tightly bound binary system (Hillebrandt, 2000). These stars have masses that do not exceed the Chandrasekhar limit (Chandrasekhar, 1931), i.e roughly 1.38 solar masses. The mass limit and the implosion mechanism are such that the characteristic light curve of an SnIa is quite uniform and reproducible, so that this kind of objects can be used as *standard candles* for determining cosmic distances (Colgate, 1979).

In order to exploit the mentioned beautiful property of SnIa's we need the *luminosity distance* of the source which depends on the expansion mechanism of the universe. When expressed in terms of *distance modulus* and of the redshift parameter it is given by the formula (Weinberg, 1972):

$$m - M = 25 + 5\log_{10}\left[(1+z)\int_0^z \frac{dz'}{H(z')}\right] \quad (28)$$

$M$ is the absolute magnitude of the source; $m$ is the locally observed magnitude; $H$ is the Hubble parameter and depends on the expansion model one uses. Formula (28) holds when distances are measured in Mpc.

When applying (28) to the luminosity data from SnIa's in the framework of the standard cosmology, one finds (Riess, 1998) (Perlmutter, 1999) that the sources appear to be dimmer than expected from the $z$ value of the host galaxy. The immediate interpretation of this fact is that the expansion of the universe is indeed accelerated.

We applied the SST to try and fit the luminosity data from SnIa's using formulae (28) and (24) (Tartaglia, 2010). The experimental luminosities were from 307 SnIa's from the *Supernova Cosmology Project Union Survey* (Kowalski, 2008). The result is shown in fig. 5; the quality of the fit, if taken as the only test, is good. The free parameters of the theory, considering that for $z$ values < 2 the radiation term is negligible, are three; the final reduced $\chi^2$ is 1.017.

For comparison we use the Λ Cold Dark Matter (ΛCDM) scenario (Concordance Model), which is the simplest and most effective theory currently adopted in order to account for the properties of the universe. ΛCDM, when employed to fit the same data of SnIa's as above, gives $\chi^2$ = 1.019. The problem with ΛCDM is that the physical nature of the cosmological constant Λ (or of the corresponding *dark energy*) remains a mystery.

For further analysis it is convenient to explicitly reproduce the $\chi^2$ formula:

$$\chi^2_{SnIa} = \sum_i \left(\frac{d_i - d(z_i)}{\delta d_i}\right)^2 \quad (29)$$

The $d_i$'s are the measured values of the distance modulus; $d(z_i)$ is the corresponding value given by the theory; $\delta l_i$ are the variances of the experimental data; the sum is over the number of supernovae we use.

This first test is encouraging, but is not enough, so let us go on with more.

## 4.2 More tests

### 4.2.1 The abundance of primordial isotopes

The lightest elements up to lithium Li[7] (mentioning just the stable isotopes) formed after the baryogenesis phase, while the primordial plasma cooled and expanded (Big Bang Nucleosynthesis: BBN). The relative abundances of hydrogen, deuterium and helium that we find today as a residue of that time depend on the early expansion history, affecting both the temperature and the density of the plasma. Since the SST gives an additional contribution to the radiation density and pressure, as seen in formula (26), we do not expect it to influence the cross section of the nuclear reactions but the quantitative final result of BBN.

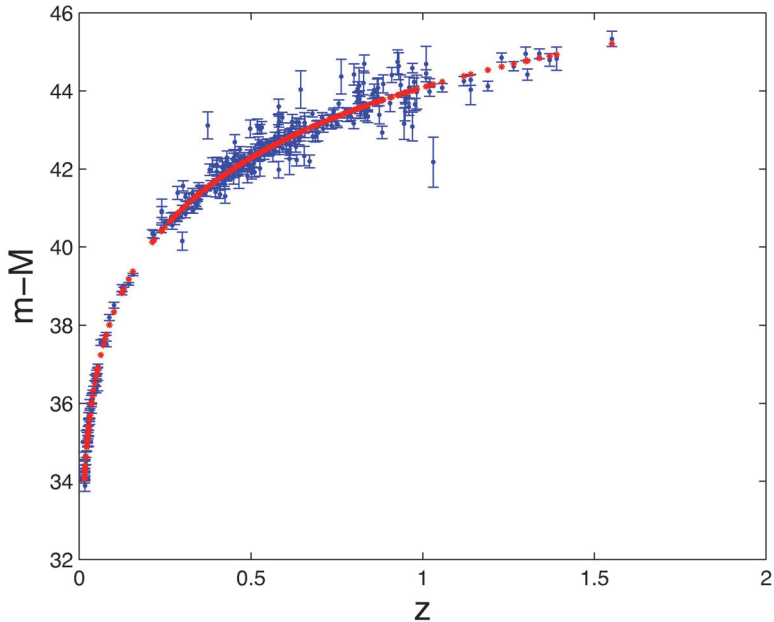

Fig. 5. Fit of the luminosity data from 307 Snia's obtained using the SST. The distance modulus is given as a function of the redshift parameter. The experimental data are shown with their error bars.

Let us recast (26) as:

$$H \cong cz^2 \sqrt{\frac{\kappa}{6}\left(1 + \frac{B}{B_{a_0}}\right)\rho_{r0}} \qquad (30)$$

where it is

$$B_{a_0} = \frac{8}{9}\kappa\rho_{r0}a_0^4 \quad (31)$$

The term in brackets in (30) acts as an effective boost factor for the radiation energy density $X_{boost} = 1 + B/B_{a0}$ enhancing the expansion rate. This fact would lead to an earlier freeze-out of the neutrons, then to a higher final abundance of He[4]. Knowing the actual abundance of helium we can then put constraints on the value of the parameters of the SST. The primordial fraction of helium by mass, $Y_p$, is estimated using various methods and with good accuracy; see for instance (Izotov, 2010). We adopted a conservative attitude picking up the value *$Y_p$ = 0.250 ± 0.03* (Iocco, 2009) obtained by an ample analysis of a number of different values in the literature. The ensuing constraint in the boost factor is *$X_{boost}$ = 1.025 ± 0.015*. Our final purpose is to perform a statistical analysis of the compatibility of SST with the data, so we work out the $\chi^2$ constraint that follows from the quoted uncertainties:

$$\chi^2_{BBN} = \left(\frac{X_{boost} - 1.025}{0.015}\right) \quad (32)$$

### 4.2.2 Cosmic microwave background constraint

The analysis of the CMB spectrum is a complex task, but the scope of this discussion is limited to a compatibility check, so I shall pick out just one parameter whose value is affected both by the expansion factor at the matter/radiation equality time and by the history of the universe from the decoupling time to the present. The chosen parameter is the acoustic scale (Komatsu, 2011):

$$l_A = (1 + z_{LS})\pi \frac{D_A(z_{LS})}{r_s(z_{LS})} \quad (33)$$

$D_A$ is the angular diameter distance to the last scattering surface; $r_s$ is the size of the sound horizon at recombination; $z_{LS} \sim 1090$ is the last scattering redshift. The mode of the expansion affects the position of the acoustic peaks which depends on the expansion factor at the equality scale $a_e$; in practice the position is influenced by the value of the boost factor for the radiation $X_{boost}$. The acoustic horizon formula will then be the same as for ΛCDM, but the equality scale factor is now boosted: $a_e = X_{boost}\rho_{r0}/\rho_{m0}$. As for the angular diameter distance, it depends on the total expansion history from the last scattering surface to present:

$$D_A(z_{LS}) = \frac{c}{(1+z_{LS})}\int_0^{z_{LS}} \frac{dz}{H(z)} \quad (34)$$

The final value for $l_A$ is not much sensitive to the choice of the cosmological model so we will make reference to the values obtained from WMAP-7 using ΛCDM (Komatsu, 2011). Our reference experimental (+ΛCDM) value is $l_A^{Obs} = 302.69 \pm 0.76 \pm 1.00$. The first uncertainty is the statistical error, the second is an estimate of the uncertainty connected with the choice of the model; the two uncertainties are mutually independent so they can be added in quadrature. Summing up we have the statistical constraint:

$$\chi^2_{CMB} = \left(\frac{l_A - 302.69}{1.26}\right)^2 \tag{35}$$

### 4.2.3 Large scale structure formation

If space-time is expanding in a radiation dominated universe matter density fluctuations cannot produce growing seeds for future structures. As we have seen, the presence of strain in early epochs effectively increases the radiation density, so retarding the onset of matter dominance. This is the reason why LSS poses further constraints on the SST. The effective boost, $X_{boost}$, affects the scale of the particle horizon at the equality epoch, $z_{eq} \cong 3150$ (Komatsu, 2011). On the other hand, the SST preserves the Newtonian limit of gravity even in presence of defects (Tartaglia, 2010), so that, in SSC, the growth of mass density perturbations is affected mainly through the modified expansion rate of the background. The horizon at the equality is imprinted in the matter transfer function. The constraint from LSS can be written as (Peacock, 1999):

$$(\Omega_{m0}h)_{apparent} = \frac{(\Omega_{m0}h)_{true}}{\sqrt{X_{boost}}} \tag{36}$$

$\Omega_{m0}$ is the mass density in units of the critical density $\rho_c = 3H_0^2/8\pi G$; $H_0$ is the Hubble constant; $h$ is the Hubble constant in units of 100 km s$^{-1}$Mpc$^{-1}$.

According to the conclusions drawn from the analysis of the data from the 2dF Galaxy Redshift Survey (Cole, 2005) it is $(\Omega_{m0}h)_{apparent} = 0.168 \pm 0.016$. For consistency we make the same assumption as in ref. (Cole, 2005) on the index of the primordial power spectrum ($n = 1$). The related constraint on the cosmological parameters of the SSC is:

$$\chi_{LSS}^2 = \left(\frac{(\Omega_{m0}h/\sqrt{X_{boost}}) - 0.168}{0.016}\right)^2 \tag{37}$$

### 4.3 Global consistency

The various tests we have described in the previous sections must be satisfied together, so we must check for the global compatibility of the constraints when applied to SSC. The analysis has been made using standard Bayesian methods (Mackay, 2003). According to Bayes theorem the posterior probability $p$ for a given parameter **P** given the data **d** is proportional to the product of the likelihood $\mathcal{L}$ of **P** times the prior probability for **P**:

$$p(\mathbf{P}|\mathbf{d}) \propto \mathcal{L}(\mathbf{P}|\mathbf{d}) p(\mathbf{P}) \tag{38}$$

The likelihood is expressed in terms of the total $\chi^2$ as $\mathcal{L} \propto e^{-\chi^2/2}$ and the total $\chi^2$ is in turn given by the sum of the independent values (29), (32), (35), (37):

$$\chi^2 = \chi^2_{Snla} + \chi^2_{BBN} + \chi^2_{CMB} + \chi^2_{LSS} \tag{39}$$

For this analysis we use three parameters of the theory. The constraints we have considered do not require us to distinguish between baryonic and dark matter, so that we consider a single parameter density for the dustlike matter, $\rho_{m0}$. The strain related properties, in a RW symmetry, are accounted for by the *B* parameter. Finally, the present value of the scale factor is described in terms of $B_{a0}$ (actually we shall use its inverse). A flat distribution for each parameter has been assumed. The relativistic energy density has been fixed at $\rho_{r0} \cong 7.8 \times 10^{-31}$ kg/m$^3$. The parameter space has been explored with Monte Carlo Markov chain methods (Lewis, 2002) running four chains, each one with $10^4$ samples. Convergence criteria were safely satisfied, with the Gelman and Rubin ratio (Gelman, 1992) being ≤ 1.003 for each parameter. The final results are shown in fig. 6a,b,c.

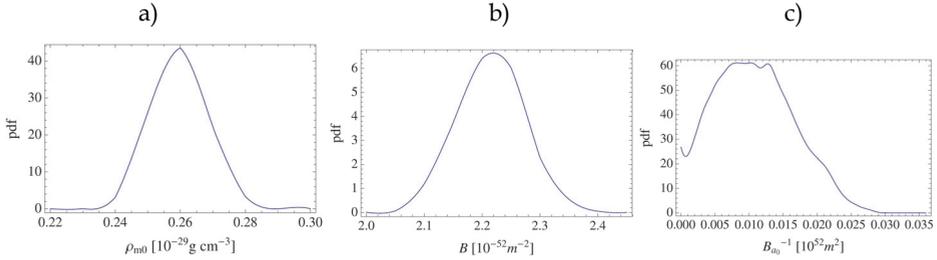

Fig. 6. Posterior probability density functions for the parameters of the SSC; the functions are normalized. Units are as in Table 1.

From the probability density functions we obtain the best estimates for the parameters. The corresponding amounts are listed in Table 1 where also the maximum likelihood values are reported in parentheses.

| $\rho_{m0}$ (10$^{-26}$ kg×m$^{-3}$) | B (10$^{-52}$ m$^{-2}$) | $B_{a0}^{-1}$ (10$^{52}$ m$^2$) |
|---|---|---|
| 0.260 (0.258) ± 0.009 | 2.22 (2.22) ± 0.06 | 0.011 (0.009) ± 0.006 |

Table 1. Estimated values of the parameters. The numbers in brackets correspond to the maximum likelihood.

The estimated value for the present matter density, when expressed in terms of the critical density, becomes $\Omega_{m0} = 0.28 \pm 0.01$ which is consistent with the value commonly accepted for the sum of baryonic and dark matter.

**4.3.1 Further compatibility checks**

The theory, together with the values obtained in the previous section for the parameters, can be used to evaluate various cosmic quantities that can be verified with observation. For instance the calculated Hubble constant of SSC is $H_0 = 70.2 \pm 0.5$ km s$^{-1}$Mpc$^{-1}$, which compares well with $73 \pm 2 \pm 4$ km s$^{-1}$Mpc$^{-1}$ obtained from high precision distance determination methods (Freedman, 2010). Another interesting quantity is the age of the universe; the SSC value is $T = 13.7$ Gy, fully compatible with the lowest limits obtained from the age of the oldest globular clusters and from radioactive dating.

## 5. Open problems and perspectives

The Strained State Theory applied to cosmology, at least in the case of a RW symmetry, performs well, as we have seen, however some aspects of the theory require further thought and clarification. Let us for instance consider a problem I have hardly touched in the previous sections: the signature of space-time. The logic of the method I have outlined here requires a totally undifferentiated flat manifold to start with. In other words the reference manifold should best be Euclidean. It is easy to verify however that the results concerning a RW universe can be obtained as well starting with a Minkowski reference manifold. The latter choice is in a sense friendlier because it has, from the start, the same signature as the final strained space-time which we want to describe. However we may ask where does the initial signature of a Minkowski space-time come from. Hopefully in the case of SSC the start can be Euclidean even if the final state has a Lorentzian signature. In the theory a cosmic defect is essential to define the global symmetry of the universe on a large scale and all timelike world lines stem out of that defect. Is the presence of a defect the condition for introducing the signature (in practice the light cones) in the natural manifold? The guess is that it is so, but the fact that the idea works in the case of the RW symmetry is not a proof, that should be given in general terms. In any case an important remark is that there must be no confusion between the reference manifold, which is Euclidean, and the local tangent space at any position in the natural manifold, which is instead Minkowskian.

The importance of the Cosmic Defect (CD) has been stressed more than once in this chapter. Are there other defects in the universe? The answer is in principle yes of course, but, if other defects exist, how and where do they show up? The CD is space-like and is the origin of the signature of space-time; if additional defects exist they could/should be time-like. A possibility is to have, for instance, a linear time-like defect; such defect would be surrounded, at any given moment, by a spherically symmetric space. If we think for instance to a big spherical cosmic void it could indeed be centered on a linear time-like defect. On the other side the present theory, for the essential, is not different from General Relativity: it is not locally distinguishable from GR, since the gravitational interaction is described in the same geometrical terms. The natural manifold admits locally a flat Minkowskian tangent space, just as in GR, and this means that the equivalence principle holds and also that the SST complies with the Newtonian limit. By the way the values obtained from the cosmological application and listed in Table 1 tell us that the scale at which deviations from the standard GR can be expected are very large, much wider than the solar system and even than a single galaxy. It is however true that the local spherical symmetry is also the typical Schwarzschild symmetry and there GR has a singular exact solution. Today black holes are well accepted and evidence for their existence, at least in the center of galaxies, is abundant. The conceptual problems posed by the singularity are bypassed by the cosmic censorship principle, so that people do not worry too much about them. Is there a connection between the black holes of GR and linear defects of the SST? The singularities of GR have to do with infinite matter densities; the defects of the SST are in the space-time as such and at most they influence the behaviour of surrounding matter. The singularity of a defect in a manifold is much friendlier than the singularities of GR. Are there horizons in SST too? All these open questions deserve further work and analysis. Remaining in the domain of defects, the properties of other symmetries need to be explored, first of all the screw symmetry which corresponds to the same symmetry as the one of the Kerr black holes.

Looking at the Lagrangian density contained in eq. (15) and in particular to the additional new elastic potential terms of eq. (14) we see that they look very much like the massive gravity Lagrangian density initially proposed by Fierz and Pauli (Fierz, 1939) (Dvali, 2008). This similarity is very strict when it is $\lambda = -2\mu$, however it must be kept in mind that the Fierz-Pauli Lagrangian was proposed in pursuit of a gravitational spin-2 field in a Minkowski background; furthermore the Fierz-Pauli Lagrangian is obtained by a linearization process in which the deviation from the flat Minkowsky manifold is represented by a $h_{\mu\nu}$ tensor, whose elements are all small with respect to 1. When letting the mass of the graviton in the Fierz-Pauli theory go to zero, one is left with a linearized General Relativity, whose equations can be used both for the study of gravito-magnetic effects and for Gravitational Waves (GW). Fierz and Pauli's approach however has a problem: its limit for zero mass of the graviton does not smoothly reproduce the results of GR: it is the so called van Dam-Veltman-Zakharov (vDVZ) discontinuity (van Dam, 1970) (Zakharov, 1970). Furthermore a non-zero mass graviton implies the presence of a ghost when studying propagating modes. The debate on these problems and on massive gravity is open.

In any case we must remark that in the SST the strain tensor is not a perturbation of a flat Minkowski background, rather it expresses the difference (not necessarily small) with respect to an Euclidean reference, which is of course *not* the tangent space at any given event of the natural manifold. The behaviour of a strained space-time with respect to propagating perturbations, i.e. waves, must be studied, but we can expect it to be similar, even though not identical, with "massive gravity"; in particular we can expect subluminal waves and contributions to a cosmic thermal gravitational background according to some appropriate dispersion law.

As a last conceptual aspect to be considered with the SST I start from a simple remark. The classical theory of elasticity is the macroscopic manifestation of an underlying microscopic reality made of discrete particles with their interactions. Can we think the elasticity of space-time to have a similar origin? The idea, at first sight, seems reasonable, however the point is subtle. On one side, an underlying microscopic structure of space-time would bring us close to the attempts to quantize the space-time and gravity (and to their difficulties). On the other, we should face the problem I mentioned in the Introduction concerning the implicit request of a "background" (a super-space-time?) in which the microscopic structure of space-time would be located. Our current view of the universe, whether we are aware of it or not, is basically dualistic: on one side space-time with properties of its own; on the other side matter/energy described by quantum mechanics in terms of eigenstates and eigenvalues of quantum operators associated with physically meaningful parameters. The two sides of the duality resist against the attempts to reduce them to a single paradigm. Maybe this simply means that nobody has found the right way so far, but it could also be that they are mutually irreducible. If so the elasticity of the four-dimensional manifold could be a fundamental property of space-time and not the macroscopic approximation of some unknown microscopic structure.

## 6. Conclusion

In this chapter I have expounded a theory based on physical intuition, which extends to four dimensions what we already know in three when studying material continua. I have used concepts such as strain to describe the distortion induced in space-time either by the presence of matter/energy or by the presence of texture defects analogous to the ones we find in crystalline solids. The idea of an induced strain implies directly the existence of an analogue of the deformation energy. This distortion energy enters the Lagrangian of space-time as an additional potential and leads to a new dynamical history of the universe. The structure and fundaments of General Relativity are all preserved. As we have seen, the theory, when applied to a Robertson-Walker universe, passes various important consistency tests, while reproducing the luminosity/distance curve of type Ia supernovae (in practice it accounts for the accelerated expansion). The values we found for the parameters of the theory tell us that locally it will be indistinguishable from GR, while producing emerging effects at cosmic scales. There are a number of developments to be pursued and difficulties to be discussed and overcome, but the way through seems not to be impassable.

Of course there are many theories that, in a way or another, account for the accelerated expansion while passing various cosmological consistency tests. First of all there is ΛCDM, which is reasonably simple and reasonably successful, though not exempt from drawbacks. How and why should we discard one theory and prefer another? Most often in cosmology new theories are introduced manipulating the Lagrangians or adding fields on heuristic bases; internal consistency is of course cared of, but physical intuition plays a minor role. Hundreds of papers appear every years discussing details of theories whose basic assumptions are motivated only by the final results one wants to obtain; the old Occam's razor (*entia non sunt multiplicanda praeter necessitatem*) is left behind and it is difficult, if not impossible, to think of crucial experiments that can discriminate among the theories. In this situation maybe the strategy of sticking as far as possible to what one already knows is sound and trying to build the least possible exotic physical scenario is advisable. This is the meaning of the Strained State Theory and of the Strained State Cosmology, which is not yet an accomplished paradigm, but aspires to become so. We have just started.

## 7. Acknowledgment

The present work could not have been carried to the level where it is now without the collaboration of a number of younger colleagues and students. I wish here to explicitly acknowledge the contributions by Ninfa Radicella and Mauro Sereno who helped me to clarify many aspects of the theory in fruitful discussions, besides co-authoring a couple of papers that have been used to support a good portion of the present chapter.